\documentclass[a4paper,12pt]{article}
\usepackage[T1]{fontenc}
\usepackage{amsfonts}
\usepackage{ae,aecompl}
\usepackage{amsmath}
\usepackage{amssymb}
\usepackage{amsthm}
\usepackage[pdftex]{graphicx}
\usepackage{pstricks}
 \usepackage{epstopdf}
\usepackage[small]{caption}   
\usepackage{color}
\usepackage{enumerate} 
\usepackage{floatrow}
\usepackage{hyperref}
\usepackage{mathdots}

\definecolor{lightgray}{gray}{0.8}

\title{Tomographic X-ray data of a walnut}
\author{K. H\"am\"al\"ainen\footnote{Department of Physics, University of Helsinki, Finland (keijo.hamalainen@helsinki.fi)},
L. Harhanen\footnote{Department of Engineering Design and Production, Aalto University, Finland (lauri.harhanen@aalto.fi)},
A. Kallonen\footnote{Department of Physics, University of Helsinki, Finland (aki.kallonen@helsinki.fi)}, \\
A. Kujanp\"a\"a\footnote{Department of Mathematics and Statistics, University of Helsinki, Finland (antti.kujanpaa@helsinki.fi)},
E. Niemi\footnote{Department of Mathematics and Statistics, University of Helsinki, Finland (esa.niemi@helsinki.fi)}  
\ and S. Siltanen\footnote{Department of Mathematics and Statistics, University of Helsinki, Finland (samuli.siltanen@helsinki.fi)}}

\begin{document}

\maketitle

\abstract{This is the documentation of the tomographic X-ray data of a walnut made available at \url{http://www.fips.fi/dataset.php}. The data can be freely used for scientific purposes with appropriate references to the data and to this document in \href{http://arxiv.org/}{arXiv}. The data set consists of (1) the X-ray sinogram of a single 2D slice of the walnut with three different resolutions and (2) the corresponding measurement matrices modeling the linear operation of the X-ray transform. Each of these sinograms was obtained from a measured 120-projection fan-beam sinogram by down-sampling and taking logarithms. The original (measured) sinogram is also provided in its original form and resolution. In addition, a larger set of 1200 projections of the same walnut was measured and a high-resolution filtered back-projection reconstruction was computed from this data; both the sinogram and the FBP reconstruction are included in the data set, the latter serving as a ground truth reconstruction.}

\section{Introduction}

The main idea behind the project was to create real CT measurement data for testing sparse-data tomography algorithms. A walnut shares some of the challenging features with the targets in typical CT applications. It contains a dense, layered shell enclosing various structures with different shapes and contrasts. Like a skull, the shell has reflection-symmetry, while the edible part of a walnut contains enough non-convexity for testing purposes.

The CT data in this data set has been used for testing a wavelet-based sparsity-promoting reconstruction method and the so-called ``S-curve method'' for choosing the regularization parameter, see \cite{Hamalainen2012a} or \cite{Mueller2012}. Moreover, the same data was used for testing the total variation reconstruction algorithms introduced and studied in \cite{Hamalainen2014}.

\section{Contents of the data set}

The data set contains the following MATLAB\footnote{MATLAB is a registered trademark of The MathWorks, Inc.} data files:
\begin{itemize}
\item  {\tt Data82.mat},
\item  {\tt Data164.mat},
\item  {\tt Data328.mat}, 
\item {\tt FullSizeSinograms.mat} and
\item  {\tt GroundTruthReconstruction.mat}.
\end{itemize}
The first three of these files contain CT sinograms and the corresponding measurement matrices with three different resolutions; the data in files {\tt Data82.mat}, {\tt Data164.mat} and {\tt Data328.mat} lead to reconstructions with resolutions $82\times 82$, $164\times 164$ and $328\times 328$, respectively. The data file named {\tt FullSizeSinograms.mat} includes the original (measured) sinograms of 120 and 1200 projections, and {\tt GroundTruthReconstruction.mat} contains a high-resolution FBP reconstruction computed from the 1200-projection sinogram. Detailed contents of each data file below.

\bigskip\noindent
{\tt Data82.mat} contains the following variables:
\begin{enumerate}
\item Sparse matrix {\tt A} of size $9840\times 6724$; measurement matrix.
\item Matrix {\tt m} of size $82\times 120$; sinogram (120 projections).
\end{enumerate}

\bigskip\noindent
{\tt Data164.mat} contains the following variables:
\begin{enumerate}
\item Sparse matrix {\tt A} of size $19680\times 26896$; measurement matrix.
\item Matrix {\tt m} of size $164\times 120$; sinogram (120 projections).
\end{enumerate}

\bigskip\noindent
{\tt Data328.mat} contains the following variables:
\begin{enumerate}
\item Sparse matrix {\tt A} of size $39360\times 107584$; measurement matrix.
\item Matrix {\tt m} of size $328\times 120$; sinogram (120 projections).
\end{enumerate}

\bigskip\noindent
{\tt FullSizeSinograms.mat} contains the following variables:
\begin{enumerate}
\item Matrix {\tt sinogram120} of size $2296\times 120$; original (measured) sinogram of 120 projections.
\item Matrix {\tt sinogram1200} of size $2296\times 1200$; original (measured) sinogram of 1200 projections.
\end{enumerate}

\bigskip\noindent
{\tt GroundTruthReconstruction.mat} contains the following variables:
\begin{enumerate}
\item Matrix {\tt FBP1200} of size $2296 \times 2296$; a high-resolution filtered back-projection reconstruction computed from the larger sinogram of 1200 projections of the walnut (``ground truth'').
\end{enumerate}

\bigskip\noindent
Details on the X-ray measurements are described in Section \ref{sec:Measurements} below.
The model for the CT problem is
\begin{equation}\label{eqn:Axm}
 {\tt A*x=m(:)},
\end{equation}
where {\tt m(:)} denotes the standard vector form of matrix {\tt m} in MATLAB and {\tt x} is the reconstruction in vector form. In other words, the reconstruction task is to find a vector {\tt x} that (approximately) satisfies \eqref{eqn:Axm} and possibly also meets some additional regularization requirements.
A demonstration of the use of the data is presented in Section \ref{sec:Demo} below.

\section{X-ray measurements}\label{sec:Measurements}

\begin{figure}
\begin{picture}(390,180)
\put(0,0){\includegraphics[width=280pt]{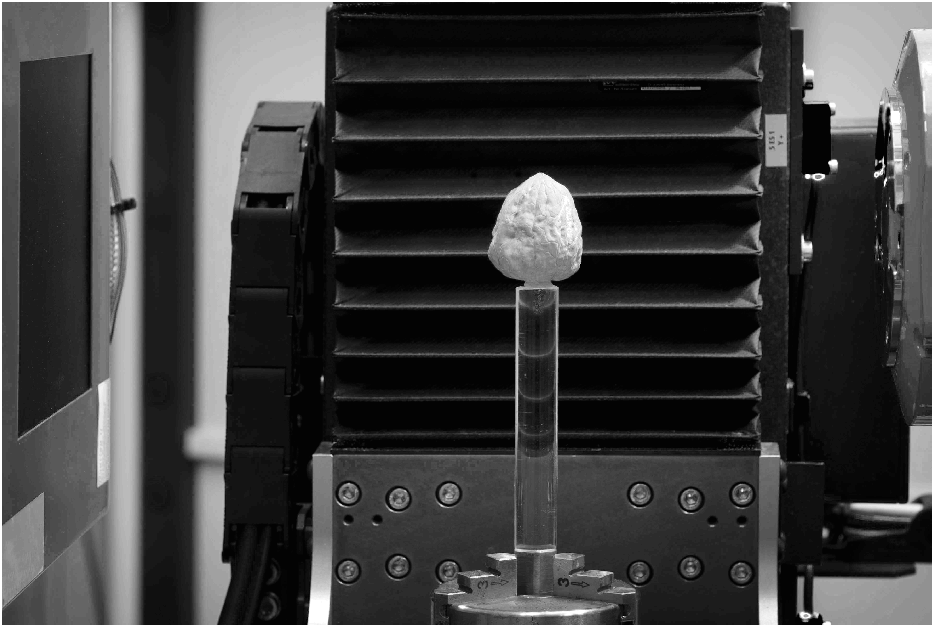}}
\put(290,0){\includegraphics[height=93pt]{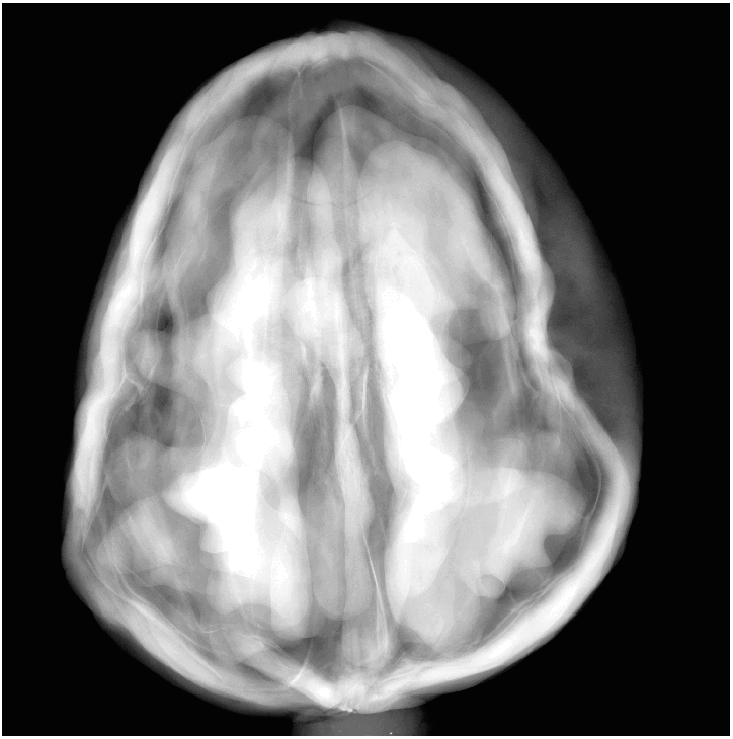}}
\put(290,95){\includegraphics[height=93pt]{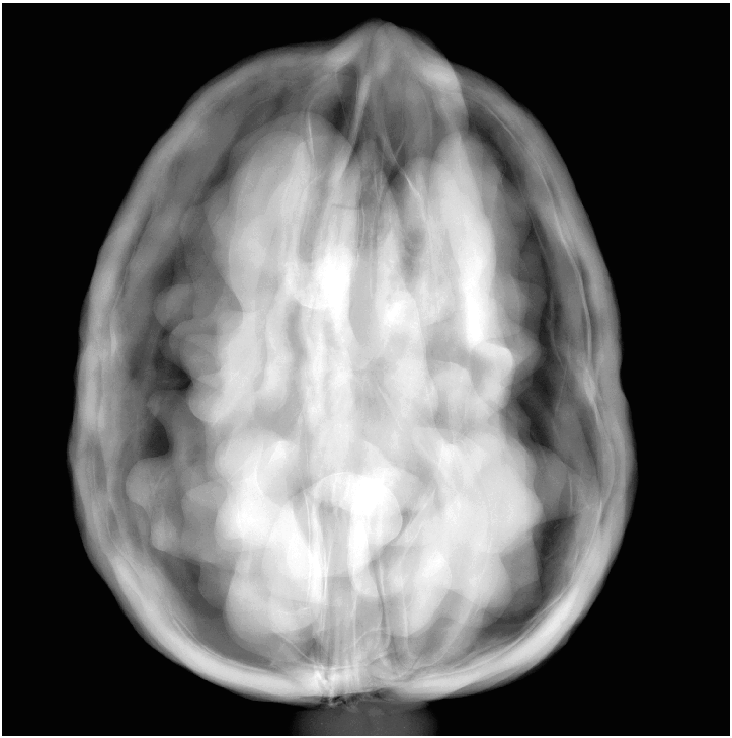}}
\put(282,46){\vector(1,0){7}}
\put(390,46){\vector(-1,0){7}}
\put(282,141){\vector(1,0){7}}
\put(390,141){\vector(-1,0){7}}
\end{picture}
\caption{\emph{Left}: Experimental setup used for collecting tomographic X-ray data of a walnut. The detector plane is on the left and the X-ray source on the right in the picture. The walnut is attached to a computer-controlled rotator platform. \emph{Right}: Two examples of the resulting 2D projection images. The fan-beam data in the sinograms consist of the (down-sampled) central rows of the 2D projection images (indicated by arrows in the picture.)}\label{fig:SetupAndProjections}
\end{figure}
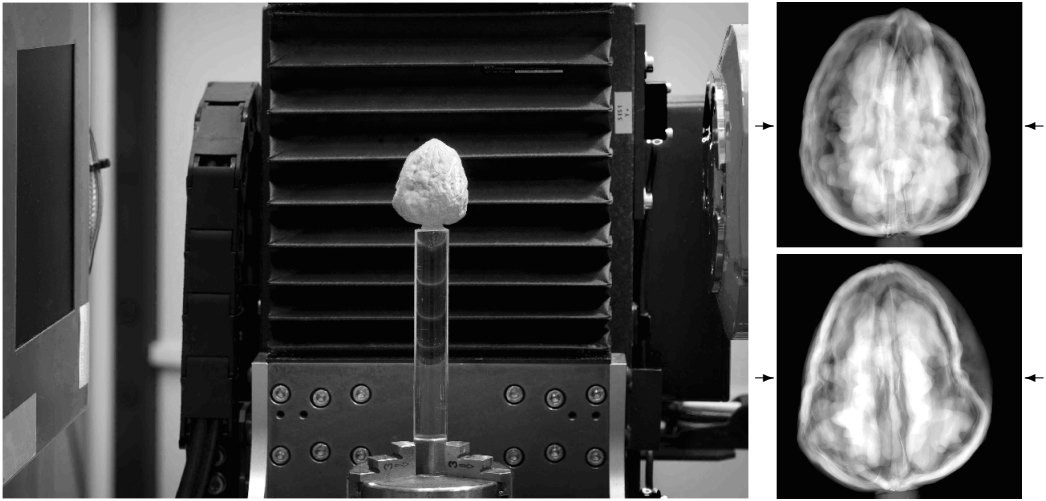

\begin{figure}
\includegraphics[width=300pt]{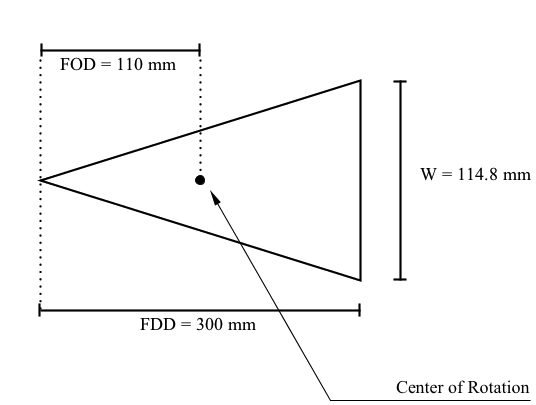}
\caption{Geometry of the measurement setup. Here FOD and FDD denote the focus-to-object distance and the focus-to-detector distance, respectively. The width of the detector is denoted by W.}\label{k1}
\end{figure}

The data  in the sinograms are X-ray tomographic (CT) data of a 2D cross-section of a walnut measured with custom-built $\mu$CT device Nanotom supplied by Phoenix|Xray Systems + Services GmbH (Wunstorf, Germany). The X-ray detector used was a CMOS flat panel detector with 2304$\times$2304 pixels of 50 $\mu$m size (Hamamatsu Photonics, Japan). The measurement system is illustrated in Figure \ref{fig:SetupAndProjections} and the measurement geometry is shown in Figure \ref{k1}. 

A  set of 120 cone-beam projections with resolution $2304\times 2296$ and angular step of three (3) degrees was measured. Each projection image was composed of an average of six 750 ms exposures. The x-ray tube acceleration voltage was 80 kV and tube current 200 $\mu$A. See Figure \ref{fig:SetupAndProjections} for two examples of the resulting projection images. From the 2D projection images the middle rows corresponding to the central horizontal cross-section of the walnut were taken to form a fan-beam sinogram of resolution $2296\times 120$ (variable {\tt sinogram120} in file {\tt FullSizeSinograms.mat}). This sinogram was further down-sampled by binning and taken logarithms of to obtain the sinograms {\tt m} in files {\tt Data82.mat}, {\tt Data164.mat} and {\tt Data328.mat}.

\begin{figure}
\begin{picture}(390,390)

\put(195,390){\lightgray\line(1,-4){90}}
\put(195,390){\lightgray\line(-1,-4){90}}

\put(145,130){\line(1,0){100}}
\put(145,155){\line(1,0){100}}
\put(145,180){\line(1,0){100}}
\put(145,205){\line(1,0){100}}
\put(145,230){\line(1,0){100}}
\put(145,130){\line(0,1){100}}
\put(170,130){\line(0,1){100}}
\put(195,130){\line(0,1){100}}
\put(220,130){\line(0,1){100}}
\put(245,130){\line(0,1){100}}
\put(152,215){$x_1$}
\put(152,190){$x_2$}
\put(155,162){$\vdots$}
\put(152,140){$x_N$}
\put(172,215){$x_{\scriptscriptstyle N+1}$}
\put(172,190){$x_{\scriptscriptstyle N+2}$}
\put(180,162){$\vdots$}
\put(174,140){$x_{2N}$}
\put(200,214){$\cdots$}
\put(200,189){$\cdots$}
\put(200,165){$\ddots$}
\put(200,139){$\cdots$}
\put(225,140){$x_{N^2}$}

\put(105,5){\line(1,0){180}}
\put(105,30){\line(1,0){180}}
\put(105,5){\line(0,1){25}}
\put(130,5){\line(0,1){25}}
\put(155,5){\line(0,1){25}}
\put(260,5){\line(0,1){25}}
\put(285,5){\line(0,1){25}}
\put(110,13){$m_1$}
\put(135,13){$m_2$}
\put(160,14){$\cdots$}
\put(241,14){$\cdots$}
\put(265,13){$m_N$}

\end{picture}
\caption{The organization of the pixels in the sinograms {\tt m}\,=\,$[m_1,m_2,\ldots,m_{120N}]^T$ and reconstructions {\tt x}\,=\,$[x_1,x_2,\ldots,x_{N^2}]^T$ in the data in {\tt Data82.mat}, {\tt Data164.mat} and {\tt Data328.mat} ($N=82,164$ or $328$). The picture shows the organization for the first projection; after that the target takes $3$ degree steps counter-clockwise (or equivalently the source and detector take $3$ degree steps clockwise) and the following columns of {\tt m} are determined in an analogous manner.}\label{fig:pixelDemo}
\end{figure}
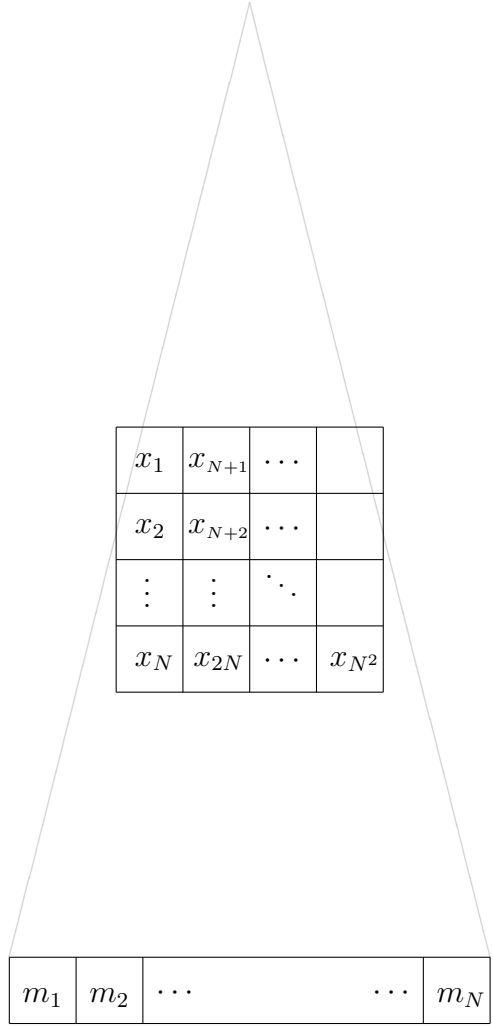

The organization of the pixels in the sinograms and reconstructions is illustrated in Figure \ref{fig:pixelDemo}. The pixel sizes of the reconstructions are $0.513$ mm, $0.257$ mm and $0.128$ mm in the data in {\tt Data82.mat}, {\tt Data164.mat} and {\tt Data328.mat}, respectively.

\begin{figure}
\includegraphics[width=\textwidth]{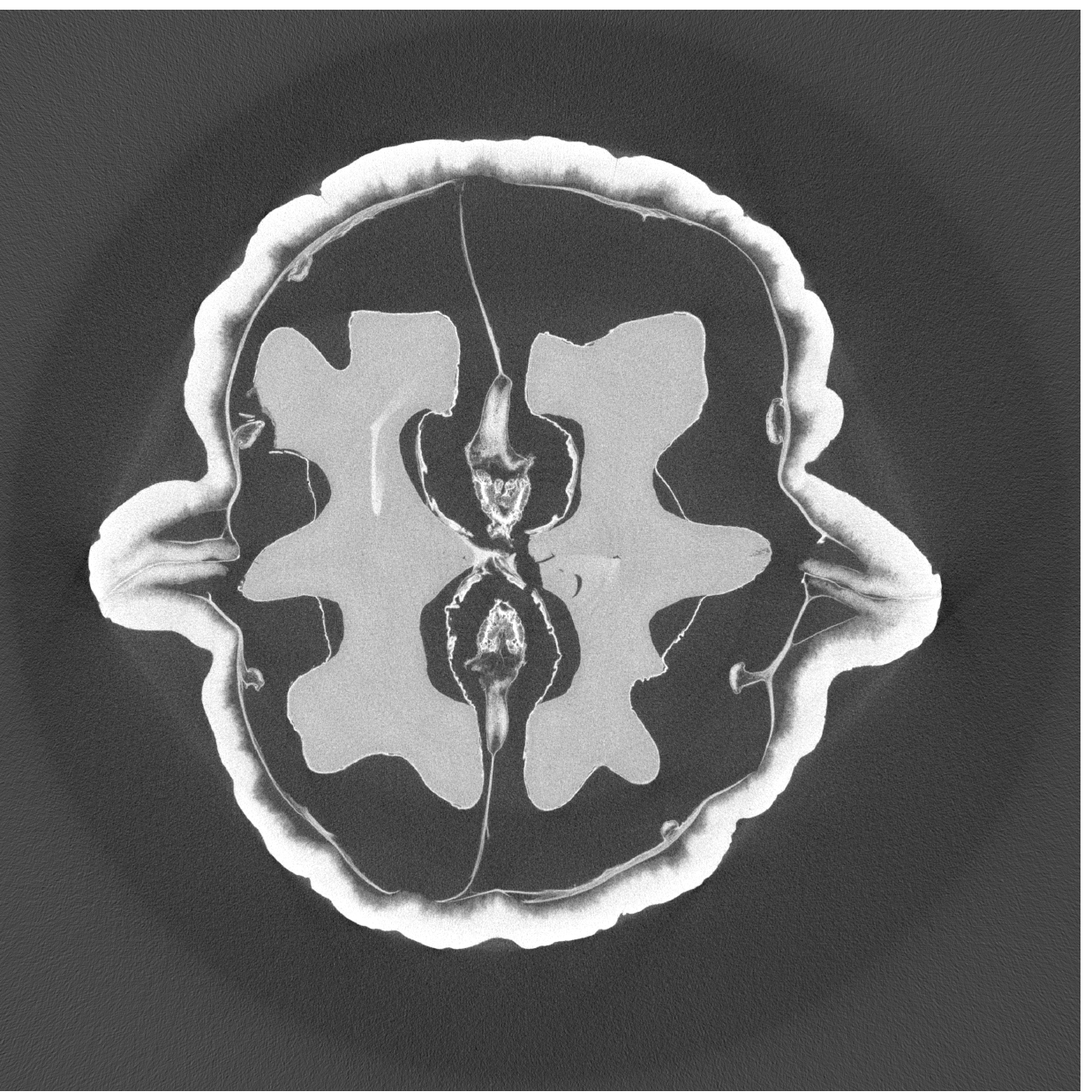}
\caption{The high-resolution filtered back-projection reconstruction {\tt FBP1200} of the walnut computed from 1200 projections.}\label{k2}
\end{figure}

In addition, a larger set of 1200 projections of the same walnut using the same imaging setup and measurement geometry but with a finer angular step of $0.3$ degrees was measured (variable {\tt sinogram1200} in file {\tt FullSizeSinograms.mat}). The high-resolution ground truth reconstruction (variable {\tt FBP1200} in file {\tt GroundTruthReconstruction.mat}) was computed from this data using filtered back-projection algorithm, see Figure \ref{k2}.

\clearpage
\section{Example of using the data}\label{sec:Demo}

The following MATLAB code demonstrates how to use the data. The code is also provided as the separate MATLAB script file {\tt example.m} and it assumes the data files (or in this case at least the file {\tt Data164.mat}) are included in the same directory with the script file.

\begin{verbatim}
% Load the measurement matrix and the sinogram from
% file Data164.mat
load Data164 A m

% Compute a Tikhonov regularized reconstruction using
% conjugate gradient algorithm pcg.m
N     = sqrt(size(A,2));
alpha = 10; % regularization parameter
fun   = @(x) A.'*(A*x)+alpha*x;
b     = A.'*m(:);
x     = pcg(fun,b);

% Compute a Tikhonov regularized reconstruction from only
% 20 projections
[mm,nn] = size(m);
ind     = [];
for iii=1:nn/6
    ind = [ind,(1:mm)+(6*iii-6)*mm];
end
m2    = m(:,1:6:end);
A     = A(ind,:);
alpha = 10; % regularization parameter
fun   = @(x) A.'*(A*x)+alpha*x;
b     = A.'*m2(:);
x2    = pcg(fun,b);

% Take a look at the sinograms and the reconstructions
figure
subplot(2,2,1)
imagesc(m)
colormap gray
axis square
axis off
title('Sinogram, 120 projections')
subplot(2,2,3)
imagesc(m2)
colormap gray
axis square
axis off
title('Sinogram, 20 projections')
subplot(2,2,2)
imagesc(reshape(x,N,N))
colormap gray
axis square
axis off
title({'Tikhonov reconstruction,'; '120 projections'})
subplot(2,2,4)
imagesc(reshape(x2,N,N))
colormap gray
axis square
axis off
title({'Tikhonov reconstruction,'; '20 projections'})
\end{verbatim}

\bibliographystyle{amsplain}
\bibliography{Inverse_problems_references}

\providecommand{\bysame}{\leavevmode\hbox to3em{\hrulefill}\thinspace}
\providecommand{\MR}{\relax\ifhmode\unskip\space\fi MR }
\providecommand{\MRhref}[2]{%
  \href{http://www.ams.org/mathscinet-getitem?mr=#1}{#2}
}
\providecommand{\href}[2]{#2}
\begin{thebibliography}{1}

\bibitem{Hamalainen2014}
K.~H\"am\"al\"ainen, L.~Harhanen, A.~Hauptmann, A.~Kallonen, E.~Niemi, and
  S.~Siltanen, \emph{Total variation regularization for large-scale x-ray
  tomography}, International Journal of Tomography and Simulation \textbf{25}
  (2014), no.~1, 1--25.

\bibitem{Hamalainen2012a}
K.~H\"am\"al\"ainen, A.~Kallonen, V.~Kolehmainen, M.~Lassas, K.~Niinim{\"a}ki,
  and S.~Siltanen, \emph{Sparse tomography}, SIAM Journal of Scientific
  Computing \textbf{35} (2013), no.~3, B644--B665.

\bibitem{Mueller2012}
J.~L. Mueller and S.~Siltanen, \emph{Linear and nonlinear inverse problems with
  practical applications}, SIAM, 2012.

\end{thebibliography}

\end{document}